\documentclass[twocolumn,showpacs,preprintnumbers,amsmath,amssymb]{revtex4}
\usepackage{graphicx}
\newcommand{\ltsim}{\protect\raisebox{-0.5ex}{$\:\stackrel{\textstyle <}{\sim}\:$}}
\newcommand{\gtsim}{\protect\raisebox{-0.5ex}{$\:\stackrel{\textstyle >}{\sim}\:$}}
\begin{document}
\title{Dynamical density matrix renormalization group study of photoexcited states in one-dimensional Mott insulators}
\author{H. Matsueda}
\email{matsueda@imr.tohoku.ac.jp}
\author{T. Tohyama}
\author{S. Maekawa}
\affiliation{Institute for Materials Research, Tohoku University, Sendai 980-8577, Japan}
\date{\today}
\begin{abstract}
One-dimensional Mott insulators exhibit giant nonlinear optical response. Based on the dynamical density matrix renormalization group method, photoexcited states and optical response in the insulators are studied as functions of the on-site and the nearest neighbor Coulomb interactions, $U$ and $V$, respectively. We find that the lowest optically allowed and forbidden excited states across the Mott gap, which have odd and even parities, respectively, are degenerate for $V/t\ltsim 2$ with $t$ being the hopping integral of an electron between nearest neighbor sites. For $V/t\gtsim 2$, the bound states with odd and even parities occur and are not degenerate. The nature of the degeneracy and its effect on the optical response are examined.
\end{abstract}
\maketitle

In one-dimensional Mott insulators with strong on-site Coulomb interaction $U$, the ultrafast and gigantic nonlinear optical response have been observed~\cite{Ogasawara,Kishida1}. The third-order nonlinear susceptibilities in the typical one-dimensional Mott insulators of charge transfer type, such as cuprates and halogen-bridged Ni-halides, are a thousand times larger than those of band insulators or Peierls insulators~\cite{Kishida1}. In the previous theoretical investigations~\cite{Kishida2,Mizuno}, it has been pointed out that the anomalous enhancement of the nonlinear susceptibilities in the one-dimensional Mott insulators originates in the large transition dipole moment between an optically allowed state with odd parity and an optically forbidden (two-photon allowed) state with even parity due to the degeneracy of these states. 

In the large-$U$ limit, only the charge degree of freedom survives in the photoexcited states due to the spin-charge separation~\cite{Ogata}, and the system is well described by an effective two particle model, called the holon-doublon model, which is composed of a holon representing one photoinduced empty site and a doublon representing one doubly occupied site with the attractive Coulomb interaction $-V$ between them. When $V$ exceeds a critical value $V=2t$, these paricles form an excitonic bound state~\cite{Stephan}. Since these two particles cannot exchange each other, the photoexcited states with odd and even parities are degenerate regardless of the magnitude of $V$. By using this model, the experimentally obtained optical spectra for the cuprates have been analyzed. However, the optical gap in the compounds is finite ($\sim 2$ eV) and thus the holon-doublon model, which is obtained in the limit of large $U$, may be re-examined. Under a finite-$U$ condition, there is an intermediate process that these particles recombine, while this process is prohibited in the large-$U$ limit. Since, in the finite-$U$ condition, these particle can exchange their positions through the intermediate process, the eigenenergies of the odd- and even-parity states may be different. Therefore, it is indispensable to clarify the optical response in Mott insulators with finite $U$. 

The minimal model to describe photoexcitation in the Mott insulators is the single-band extended Hubbard model with $U$ and $V$ at half filling. By using the model with parameters for Sr${}_{2}$CuO${}_{3}$, the linear absorption and two photon absorption spectra have been calculated in the exact diagonalization method for a 12-site chain~\cite{Mizuno}. To understand the degenerate feature in more detail, 
systematic studies of the spectra are required for large size systems by changing the parameters widely. To achieve this purpose, we employ the dynamical density matrix renormalization group (DMRG) method~\cite{Kuhner,Jeckelmann1,Jeckelmann2,Jeckelmann3,Essler}, and calculate the dynamical correlation functions for odd- and even-parity states. The finite size scaling of the lowest-energy positions, which are derived from the corrections functions, is performed. We clarify a parameter region where the degenerate photoexcitated states appear.

The single-band extended Hubbard model in one-dimension is given by
\begin{eqnarray}
H&=&-t\sum_{i,\sigma}( c^{\dagger}_{i,\sigma}c_{i+1,\sigma}+{\rm H.c.} )+U\sum_{i}n_{i,\uparrow}n_{i,\downarrow} \nonumber \\
&& +V\sum_{i}n_{i}n_{i+1} ,
\end{eqnarray}
where $c^{\dagger}_{i,\sigma}$ is the creation operator for an electron with spin $\sigma$ at site $i$, $n_{i}=n_{i,\uparrow}+n_{i,\downarrow}$, $t$ is the hopping integral along the chain axis, $U$ is the on-site Coulomb interaction, and $V$ is the Coulomb interaction between nearest neighbor sites. We consider the half-filled case. We examine the wide parameter region $U/t=5$, $10$, $20$, and $40$ to see the nature of the excited states, although it is known that the realistic value of $U/t$ for the cuprates is $U/t\sim 10$.

We focus on the low energy region of the optical spectra with even and odd parities at zero temperature. The quantity $\chi_{A}(\omega)$ for an operator $A$ is defined by
\begin{equation}
\chi_{A}(\omega)=-\frac{1}{L}{\rm Im}<0|A\frac{1}{\omega+E_{0}-H+i\gamma}A^{\dagger}|0>,
\end{equation}
where $L$ is the system size, $|0>$ is the ground state, $E_{0}$ is its energy, and $\gamma$ is an infinitesimal positive number. Opeator $A$ is either the paramagnetic current operator $j=-it\sum_{i,\sigma}( c^{\dagger}_{i,\sigma}c_{i+1,\sigma}-{\rm H.c.} )$ or the stress-tensor operator $\tau=-t\sum_{i,\sigma}( c^{\dagger}_{i,\sigma}c_{i+1,\sigma}+{\rm H.c.} )$~\cite{Shastry}. Since $j$ and $\tau$ have odd and even parities under the space inversion, respectively, $\chi_{j}(\omega)$ detects the odd-parity states while $\chi_{\tau}(\omega)$ does the even-parity states.

To calculate these quantities, we use the dynamical DMRG method~\cite{Jeckelmann2,Jeckelmann4}, which is known to be powerful to study dynamical properties in one-dimensional correlated electron systems. The basic DMRG algorithm~\cite{White} for generating the ground state restricts the exponentially increasing Hilbert space to the relevant low energy DMRG basis using the ground state density matrix. To make informations about the photoexcited states reflect on the next DMRG step, the density matrix for a system block is expanded to a following form~\cite{Kuhner,Hallberg}~:
\begin{equation}
\rho_{ii^{\prime}}=\sum_{\alpha}p_{\alpha}\left[ \sum_{j}\psi^{\ast}_{\alpha,ij}\psi_{\alpha,i^{\prime}j}/\sum_{i,j}\psi^{\ast}_{\alpha,ij}\psi_{\alpha,ij}\right] ,
\end{equation}
where $\sum_{\alpha}p_{\alpha}=1$ with $\alpha$ denoting each target state, the indices $i$ and $j$ run over all bases of a system block and an environment block, respectively. We adopt $\psi_{\alpha=0}=|0>$, $\psi_{\alpha=1}=A^{\dagger}|0>$, and $\psi_{\alpha=2}=(\omega+E_{0}-H+i\gamma)^{-1}A^{\dagger}|0>$. 
The most relevant states to go onto the next DMRG step are constructed by the eigenstates of this enlarged matrix, and the spectral functions are directly given by the imaginary part of the inner product between $\psi_{1}$ and $\psi_{2}$ after the sweep procedure is converged. We take $p_{0}=0.5$, and $p_{1}=p_{2}=0.25$. It is noted that the precise variational principle for the ground state wave function is necessarily violated whenever we treat the dynamical effects by choosing $p_{0}\neq 1$. However, we can confirm that spectral shapes evaluated by DMRG are quantitatively consistent with results from the Lanczos diagonalization for small clusters. Since we empirically know that the numerical results are generally insensitive to how to determine the parameter set $\{p_{\alpha}\}$, we believe that convergency of the obtained ground state is considerably good. 

In the numerical simulation, an edge effect by the open boundary condition appears as a side peak located below a main excitonic peak. This indicates that a doubly occupied site tends to stay near the boundary. When we additionaly introduce an impurity potential as $\tilde{H}=H+V(n_{1}+n_{L})$, the side peak is removed. We have confirmed that the spectra obtained in the method is similar to those in the exact diagonalization method for small clusters with the peroidic boundary condition. ~\cite{Tsutsui}

\begin{figure}
\begin{center}
\includegraphics[width=11cm]{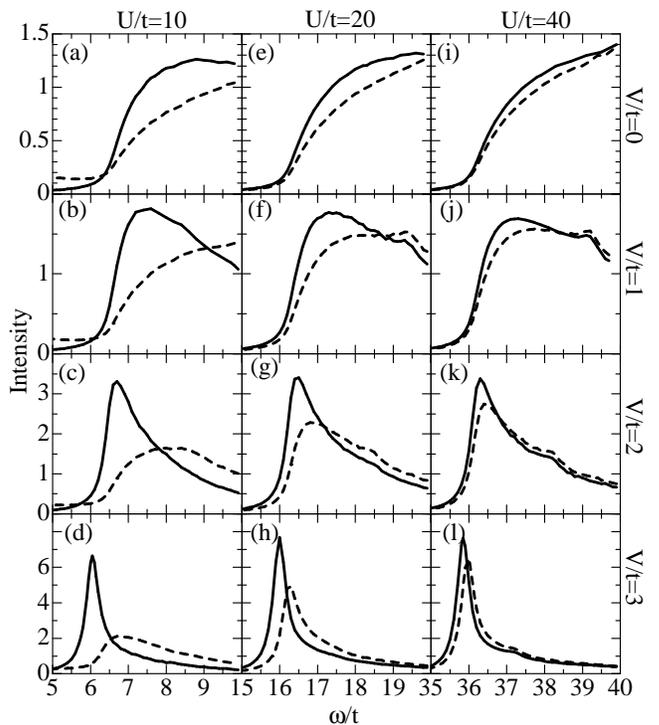}
\end{center}
\caption{$\chi_{j}(\omega)$(solid lines) and $\chi_{\tau}(\omega)$(broken lines) as functions of $U$ and $V$. The system size $L$ and the broadening factor $\gamma$ are taken to be $L=64$ and $\gamma L/t=12.8$.}
\end{figure}

We first report the dynamical correlation functions, $\chi_{j}(\omega)$ and $\chi_{\tau}(\omega)$, for several values of $U/t$ and $V/t$ in Fig.~1. The truncation number $m$ is $m=180$ in most cases of the simulation and sometimes $m=240$ if necessary. The average discarded weight is about the order of $10^{-6}$. In the largest-$U/t$ case ($U/t=40$), the spectra with odd and even parities are similar each other for $V/t\le 3$. With further increasing $U/t$ at a fixed value of $V/t$, we can expect $\chi_{j}(\omega)=\chi_{\tau}(\omega)$ for any $\omega$, since the model becomes equivarent to the holon-doublon model. In Figs.~1(i)-1(l), a tiny bump structure is seen around $\omega=U-V$, which is associated with a uniform exciton in the case of a dimerized spin background~\cite{Gebhard}.

When the parameter $U/t$ decreases at a fixed value of $V/t$, the odd-parity spectra, $\chi_{j}(\omega)$, does not change much. On the other hand, the spectral weight near the band edge in the even-parity spectra, $\chi_{\tau}(\omega)$, decreases, compared with that of $\chi_{j}(\omega)$. This is especially strong for $V/t\gtsim 2$ [see Fig.~1(d)]~\cite{Kancharla}. As already mentioned in the introduction, the excitonic bound state is formed for $V/t\ge 2$ in the large-$U$ limit. Even in finite-$U$ cases, we expect that the bound state occurs as seen in the sharp peak structure of $\chi_{j}(\omega)$ for $V/t\gtsim 2$ in Fig.~1. 

To evaluate the band edges of $\chi_{j}(\omega)$ and $\chi_{\tau}(\omega)$ more precisely, we carry out the scaling analysis in the case that $U/t=10$. 
The band edge is defined by the lowest-energy peak position evaluated using a sufficiently small broadening factor $\gamma$. The scaling results are shown in Fig.~2. For $64$-site systems, we take $m=300$. For $V/t\le 2$, it is found that the peak positions of $\chi_{j}(\omega)$ and $\chi_{\tau}(\omega)$ are almost degenerate at $L=64$. The peak positions approach to the exact Bethe ansatz solution of the band edge at $V/t=0$, which is shown by the broken line in Fig. 2~\cite{Woynarovitch}. Since the extrapolated positions in $\chi_{j}(\omega)$ and $\chi_{\tau}(\omega)$ are insensitive to the magnitude of $V/t$ for $V/t<2$, we conclude the positions to be at the continuum band edges. On the other hand, the excitonic bound states are formed below the continum band edge with a certain gap for $V/t>2$. In the case that $V/t=3$, both parity states have the excitonic feature below the continuum band edge. However, the spectral weights around the bound states behave differently as seen in Fig.~1(d). Figure~3 shows the scaling result at $U/t=5$. In this relatively small-$U$ case, the critical value of $V/t$, where the degeneracy is lifted, still seems to exist near $V/t=2$ but slightly smaller than that of the large-$U$ limit. Consequently, we conclude that in the finite-$U$ cases, the excitonic bound state formation for $V/t\gtsim 2$ lifts the degeneracy of the lowest-energy peaks in $\chi_{j}(\omega)$ and $\chi_{\tau}(\omega)$ seen for $V/t\ltsim 2$.

\begin{figure}
\begin{center}
\includegraphics[width=10cm]{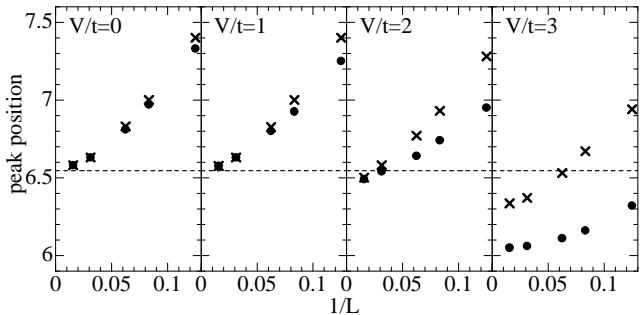}
\end{center}
\caption{Finite size scaling for the lowest-energy peak positions of $\chi_{j}(\omega)$ and $\chi_{\tau}(\omega)$ for $U/t=10$ . The filled circles denote the positions of $\chi_{j}(\omega)$, and the cross symbols denote those of $\chi_{\tau}(\omega)$. The magnitude of error bars is smaller than that of the symbols. The band edge at $V/t=0$ obtained by the Bethe ansatz method is denoted by the horizontal broken line.}
\end{figure}

The reason why the excitonic bound states are not degenerate in the case of finite $U$ is interpreted as follows:~In the holon-doublon model, there is the local constraint that double occupancy of a holon and a doublon is prohibited on the same site. On the other hand, in the case of finite $U$, there is an intermediate process that these particles recombine together. These particles can exchange their positions through the intermediate process. In such a case, nonzero matrix elements of the order of $t^{2}/U$ appear between photoexcited states before and after the exchange, leading to the lift of the degeneracy. For $V/t\gtsim 2$, a probability of the exchange is large, since these particles approach each other to form the bound states. Therefore, the lift of the degeneracy becomes remarkable for $V/t\gtsim 2$. On the other hand, a probability of finding these particles nearby is expected to be small for $V/t\ltsim 2$, since the excitonic bound states are not formed. This indicates that the effect of the exchange is negligible. This idea is supported by the DMRG results for $V/t\ltsim 2$, where the even- and odd-parity states are degenerate. We note that the behavior of the exciton for $V/t\gtsim 2$ is similar to that of a Mott-Wannier type exciton~\cite{Abe}, where an electron and a hole can exchange their positions freely. 

\begin{figure}
\begin{center}
\includegraphics[width=10cm]{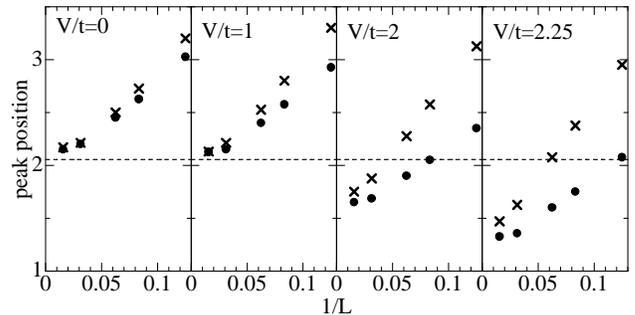}
\end{center}
\caption{Same as Fig.~2 but $U/t=5$.}
\end{figure}

Returning to the extended Hubbard model, we clarify the mechanism discussed in the previous paragraph. To simplify the discussion, a strong-coupling perturbation theory from the large-$U$ limit is applied to introduce an effective Hamiltonian. A unitary transformation to remove  the electron motion from a singly (doubly) occupied state to a doubly (singly) occupied state leads to 
\begin{eqnarray}
H&=&(U+J)\sum_{i}n_{i,\uparrow}n_{i,\downarrow}+(V-\frac{1}{4}J)\sum_{i}n_{i}n_{i+1} \nonumber \\
 && -t\sum_{i,\sigma}( \xi^{\dagger}_{i,\sigma}\xi_{i+1,\sigma} + \eta^{\dagger}_{i,\sigma}\eta_{i+1,\sigma} + {\rm H.c.} ) \nonumber \\
 && +J\sum_{i}\vec{S}_{i}\cdot\vec{S}_{i+1} \nonumber \\
 && +\frac{1}{2}J\sum_{i}( c_{i,\downarrow}^{\dagger}c_{i,\uparrow}^{\dagger}c_{i+1,\uparrow}c_{i+1,\downarrow} + {\rm H.c.} ) , \label{trans}
\end{eqnarray}
where the three site hopping terms within the same Hubbard bands are neglected, $J=4t^{2}/(U-V)$, $\xi_{i,\sigma}$ is the annihilation operator of the singly occupied state $\xi_{i,\sigma}=c_{i,\sigma}(1-n_{i,-\sigma})$, $\eta_{i,\sigma}$ is the annihilation operator of the doubly occupied state $\eta_{i,\sigma}=c_{i,\sigma}n_{i,-\sigma}$ and $\vec{S}_{i}$ is the spin operator. If we assume that all sites are occupied by one electron in the ground state, the photoexcited states created by the current or the stress-tensor are written as $(\phi_{1}\pm\phi_{2})^{\dagger}|0>$ with $\phi^{\dagger}=(\phi_{1}^{\dagger},\phi_{2}^{\dagger})=(\sum_{i,\sigma}\eta^{\dagger}_{i+1,\sigma}\xi_{i,\sigma} , \sum_{i,\sigma}\eta^{\dagger}_{i,\sigma}\xi_{i+1,\sigma})$. We evaluate a propagator matrix $D(\omega+i\gamma)=t^{2}<0|\phi(\omega+E_{0}-H+i\gamma)^{-1}\phi^{\dagger}|0>$. Considering a sign factor by fermion exchange, we can regard $-{\rm Im}(D_{11}-D_{12}-D_{21}+D_{22})/L$ as $\chi_{j}(\omega)$ and $-{\rm Im}(D_{11}+D_{12}+D_{21}+D_{22})/L$ as $\chi_{\tau}(\omega)$. By use of the standard equation of motion method for the inverse Fourier transform of $D$, the optical spectra are finally given by
\begin{equation}
\chi_{A}(\omega)=-{\rm Im}\frac{4t^{2}}{\omega+i\gamma-\omega_{A}-X_{A}(\omega+i\gamma)} ,
\end{equation}
where $A$ is either $j$ or $\tau$, $\omega_{j}=\omega_{0}-J/2$, $\omega_{\tau}=\omega_{0}+J/2$, $\omega_{0}=U-V+2J$, and $\sum_{i}<0|\vec{S}_{i}\cdot\vec{S}_{i+1}|0>/L=-3/4$. The dynamical corrections are rewritten by $X_{j}=(Y_{11}-Y_{12})/2L$, and $X_{\tau}=(Y_{11}+Y_{12})/2L$, where $Y=<0|\delta\phi(\omega+E_{0}-H+i\gamma)^{-1}\delta\phi^{\dagger}|0>_{I}$ with $I$ denoting the irreducible part of $\phi$. In a case that up and down spins on sites are alternately arranged by the interaction, $J$, $\delta\phi^{\dagger}$ is defined by $\delta\phi^{\dagger}=[2\sum_{i,s,s^{\prime}}\xi^{\dagger}_{i-1,s}(\vec{\sigma})_{ss^{\prime}}\eta_{i+1,s^{\prime}}\cdot\vec{S}_{i},2\sum_{i,s,s^{\prime}}\xi^{\dagger}_{i+1,s}(\vec{\sigma})_{ss^{\prime}}\eta_{i-1,s^{\prime}}\cdot\vec{S}_{i}]=(\sum_{i}\delta\phi^{\dagger}_{1i},\sum_{i}\delta\phi^{\dagger}_{2i})$. For particular sites, $i-1$, $i$, $i+1$, a transition process by $\delta\phi^{\dagger}_{1i}$ is given by $\delta\phi_{1i}^{\dagger}|\cdot\cdot\cdot,\uparrow,\downarrow,\uparrow,\cdot\cdot\cdot>=|\cdot\cdot\cdot,0,\uparrow,\downarrow\uparrow,\cdot\cdot\cdot>$, which means that a holon-doublon pair extends spatially. Then, $X_{A}$ corresponds to continuum band. Obviously, $X_{j}$ and $X_{\tau}$ are degenerate at band edges according to their definitions. When the edges are located below $\omega_{j}$ and $\omega_{\tau}$, the band edges of $\chi_{j}$ and $\chi_{\tau}$ are also degenerate, since hybridization between the edge of $X_{A}$ and a pole $\omega_{A}$ is negligible. We note that negligibly small hybridization corresponds to negligibly small probability of finding a holon and a doublon nearby for $V/t\ltsim 2$ discussed in the previous paragraph. When the excitonic bound states are separated from the edges, we can apply the single-pole approximation to the calculation of $X_{A}$, and obtain $X_{j}(\omega)=X_{\tau}(\omega)=4t^{2}/[\omega-(\omega_{0}+V)]$, where we assume $\sum_{i}<0|\vec{S}_{i}\cdot\vec{S}_{i+2}|0>/L=1/4$. In this case, the energy difference between odd- and even-parity states is given by 
$\Delta\omega=[\omega_{\tau}+X_{\tau}(\omega_{\tau})]-[\omega_{j}+X_{j}(\omega_{j})]=J[V^{2}-(J/2)^{2}-4t^{2}]/[V^{2}-(J/2)^{2}]$. We note that the energy difference arises from the last term of the transformed Hamiltonian (\ref{trans}) that describes exchange between a holon and a doublon. The values of $\Delta\omega$ are evaluated to be $\Delta\omega=0.315$ and $0.171$ for $(U/t,V/t)=(10,3)$ and $(5,2.25)$, respectively. These values are comparable with the DMRG results for 64-site systems shown in Figs.~2 and~3, which give $\Delta\omega=0.285$ and $0.145$, respectively. Since the exciton with even parity is situated near the continuum band due to $\Delta\omega>0$, it mixes strongly the comtinuum band. Then, the excitonic spectral weight for the even-parity state is reduced.

Among the related compounds, we focus on Sr${}_{2}$CuO${}_{3}$ to make a comparision between the present results and experiments. Mapping the Zhang-Rice band onto the lower Hubbard band, the parameters in eq. (1) for Sr${}_{2}$CuO${}_{3}$ have been estimated to be $U/t=10$ and $V/t=1.5\sim 2.0$~\cite{Mizuno,Neudert}. The parameters correspond to those used in Fig.~1(c). We find that $\chi_{j}(\omega)$ representing the distribution of the odd-parity states reproduces the measured linear absorption spectrum~\cite{Ogasawara,Kishida1}. In the case that $U/t=10$ and $V/t=2$, the absorption band edges in $\chi_{j}(\omega)$ and $\chi_{\tau}(\omega)$ are found to be nearly degenerate. The distribution of the even-parity states for Sr${}_{2}$CuO${}_{3}$ has been measured by the two-photon absorption (TPA) spectroscopy, and a finite energy difference ($\sim 0.3$ eV) between the peaks in the linear absorption and TPA spectra has been reported~\cite{Ogasawara}. The behavior of the band edges in the odd- and even-parity spectra is different from that of the peaks in the spectra. The calculation of the TPA spectrum in the present model will make it clear whether the peak position in the TPA spectrum is higher in energy than the band edge position in $\chi_{\tau}(\omega)$. Such a calculation is in progress.

Among the halogen-bridged Ni-halides, the Br-based compounds show anomalous behaviors:~the linear absorption peak is very sharp but the excitonic effect is found to be small from photoconductivity measurements although the effect can not be neglected~\cite{Okamoto}. In the present model, the optical responses observed in the compounds have not been 
obtained. Since the optical gap $\Delta$ in the compounds has been reported to be comparable with the $p-d$ transfer energy $t_{pd}$~\cite{Fujimori}, it is not clear that the single-band model describes the electronic states in the compounds. It might be necessary to study the underlying electronic states more detail.

In summary, we have presented the optical spectra with odd and even parities by changing widely a set of parameters in the one-dimensional single-band model by using the dynamical DMRG method. We have obtained three results:~(1) The excitonic bound states are formed for $V/t\gtsim 2$ in finite-$U$ cases. (2) When the excitons are unbound for $V/t\ltsim 2$, the photoexcited states with odd and even parities are degenerate. (3) When the excitonic bound states are formed, the eigenenergies of the odd and even states split because the empty site and the doubly occupied site can exchange each other in the finite-$U$ cases.

The authors thank K. Tsutsui for several suggestions for the exact diagonalization method.
This work was supported by NAREGI Nanoscience Project and Grant-in-Aid for Scientific Research from the Ministry of Education, Culture, Sports, Science and Technology of Japan, and CREST. The numerical calculations were performed in the supercomputing facilities in ISSP, University of Tokyo and IMR, Tohoku University.

\end{document}